\documentclass[
  aps,
  prl,
  reprint,
  amsmath,
  amssymb,
  footinbib,
  longbibliography
]{revtex4-2}

\usepackage{graphicx}
\usepackage[usenames,dvipsnames]{xcolor}
\usepackage{bm}
\usepackage{threeparttable}
\usepackage{fontawesome}
\usepackage[caption=false]{subfig}

\usepackage{capt-of}

\definecolor{rosewood}{rgb}{0.4, 0.0, 0.04}
\definecolor{pyblue}{RGB}{31, 119, 180}
\definecolor{pyorange}{RGB}{255, 127, 14}
\definecolor{pygreen}{RGB}{44, 160, 44}
\definecolor{pyred}{RGB}{214, 39, 40}
\definecolor{lightgray}{gray}{0.9}
\colorlet{lightpyblue}{pyblue!30!white}
\colorlet{lightpyred}{pyred!30!white}
\colorlet{lightpygreen}{pygreen!30!white}

\usepackage{hyperref}
\hypersetup{
    pdfencoding=unicode,
    colorlinks=true,
    urlcolor=rosewood,
    linkcolor=pyblue,
    citecolor=pygreen,
    pdftitle={Carving out the Multifield Cosmological Collider Landscape},
    pdfauthor={Denis Werth},
    pdfdisplaydoctitle=true,
    pdfstartview=FitH,
    linktocpage=true
}

\def \d {\mathrm{d}}
\def \x {\bm{x}}

\def \B {\mathcal{B}}

\def \H {\mathcal{H}}

\def \K {\mathcal{K}}
\def \L {\mathcal{L}}
\def \M {\mathcal{M}}
\def \N {\mathcal{N}}
\def \O {\mathcal{O}}
\def \P {\mathcal{P}}

\def \U {\mathcal{U}}

\def \SO {\mathrm{SO}}

\def \Im {\mathrm{Im}}
\def \Re {\mathrm{Re}}
\def \fnl {f_{\rm NL}}
\def \Mpl {M_{\rm pl}}
\def \meff {m_{\rm eff}}
\def \diag {\mathrm{diag}}
\def \Nf {N_{\rm f}}

\begin{document}
\hbadness=10000

\title{Carving out the Multifield Cosmological Collider Landscape}

\author{Denis Werth}
\email{werth@mpp.mpg.de}
\affiliation{Max Planck Institute for Physics, Werner-Heisenberg-Institut, Munich, D-85748, Germany,}
\affiliation{Max Planck-IAS-NTU Center for Particle Physics, Cosmology and Geometry}

%\date{\today}

%-------------------------------------
\begin{abstract}
We show that inflation with a large number of fields---as predicted by concrete ultraviolet embeddings---generically produces almost-local cosmological collider signals in the bispectrum squeezed limit. We construct the most general boost-breaking mixings between the curvature perturbation and $\Nf$ additional fields, and extract the late-time scaling dimensions that dictate the soft limits of cosmological correlators. For a handful of fields, an enhanced squeezed limit (with potential oscillations) is a fine-tuned signal. At large $\Nf=\O(100)$, however, mixings generically drive the scaling dimensions toward the unitarity boundary. We show that this is a direct consequence of extreme value statistics and random-matrix universality, turning an almost-local bispectrum shape into a generic prediction for almost any masses and mixings. 
\end{abstract}

\maketitle

%-------------------------------------
{\bf Introduction.---}Inflation, when embedded in a concrete ultraviolet (UV) completion, naturally comes with a plethora of scalar fields, including moduli in string compactifications and towers of Kaluza--Klein or axion-like states~\cite{Alishahiha:2004eh, Silverstein:2003hf, Kim:2004rp, Dimopoulos:2005ac, Tolley:2009fg, Langlois:2008mn, Langlois:2008wt}, see~\cite{Baumann:2014nda} for a review. The decay of these additional species into inflaton fluctuations leaves distinctive imprints in the soft limits of cosmological correlators, giving rise to characteristic cosmological collider signals~\cite{Chen:2009zp, Noumi:2012vr, Arkani-Hamed:2015bza, Lee:2016vti}. This opens a window onto fundamental particles and their interactions at the inflationary scale and, ultimately, onto the spectrum of the multifield sector. A central challenge is therefore to connect these theoretical predictions to cosmological observations~\cite{Sefusatti:2012ye, Meerburg:2016zdz, Cabass:2024wob, Sohn:2024xzd, Suman:2025tpv, Kumar:2026dih, Philcox:2026rpn, Philcox:2026tjj}, and identify signatures that can be searched for in~data.

The robustness of collider signals stems from their universality, dictated by symmetry. At late time during inflation, the dynamics of a field $\sigma(\tau, \x)$ is entirely fixed by its scaling dimension,
\begin{equation}
    \sigma(\tau, \x) \sim \bar\sigma_+(\x) (-\tau)^{\Delta_+} + \bar\sigma_-(\x) (-\tau)^{\Delta_-} \,.
\end{equation}
At the level of the bispectrum shape, these signals live in the squeezed limit:
\begin{equation}
\label{eq: squeezed shape}
    S \sim \sum_\pm c_\pm\left(\frac{k_\ell}{k_s}\right)^{\Delta_\pm-1} + \text{c.c.} \,, \quad (k_\ell\ll k_s) \,,
\end{equation}
where a long mode ($k_\ell$) exits the horizon much before short modes ($k_s$). Assuming de Sitter (dS) isometries, the common lore is that the principal series (associated with heavy fields in Hubble units, $\Delta_\pm\equiv r\pm i\mu$ with $r=3/2$ and $\mu\in\mathbb{R}$) gives a \emph{suppressed oscillatory} signal with frequency set by $\mu$, while the complementary series (associated with light fields, $\Delta_\pm\equiv r\pm\nu$ with $r=3/2$ and $\nu\in[-\tfrac32, \tfrac32]$) gives an \emph{enhanced power-law} signal. At the unitarity boundary $\Delta_-=0$ ($\Delta_+=3$) this reduces to the \emph{local shape}, $S\sim k_s/k_\ell$.

Yet inflation must eventually end, and the resulting breaking of dS symmetry allows fields to mix non-trivially already at the linear level. With fewer symmetries comes greater freedom. Physical propagating modes do not necessarily align with bare fields, which modifies their scaling dimensions that ultimately govern observable signals. The canonical example is the ``turn-induced'' mixing between the inflaton fluctuation and a single additional field, which dresses its mass~\cite{Chen:2009zp, Iyer:2017qzw, An:2017hlx, Pinol:2023oux, Huenupi:2026abj, Huenupi:2026aqc, Wang:2026lff, Pinol:2026xnl, Belrhali:2026uxn}. But what happens if we let the number of fields, $\Nf$, and the mixings between them be as general as the reduced symmetries allow? And what emerges when $\Nf$ is taken \emph{large}, of order $\Nf=\O(100)$, as suggested by the string landscape?

In this \emph{Letter}, we open a large-$\Nf$ window onto cosmological collider physics. Adopting a statistical random-matrix approach to scan boost-breaking mixings in multifield inflation~\cite{Tegmark:2004qd, Easther:2005zr, Frazer:2011tg, Frazer:2011br, McAllister:2012am, Masoumi:2016eag, Bjorkmo:2017nzd, Chen:2026qbn}, we show that mixings can profoundly reshape collider observables, altering the interpretation of their measured signatures. 

For a modest number of additional fields, $\Nf=\O(1)$, we find that enhanced power-law collider signals are fine-tuned signatures. As $\Nf$ becomes large, however, this picture is reversed. Enhanced, almost-local power-law signals emerge as generic predictions, as a consequence of extreme-value statistics and the universality of random-matrix ensembles: mixings drive the scaling dimensions toward the unitarity boundary. This effect disappears when all bare masses are sufficiently heavy, for which the usual cosmological-collider picture is recovered. More broadly, these results extend the cosmological-collider framework across a much wider class of multifield inflationary scenarios, helping to reduce model-dependent bias in the interpretation of cosmological data.
%Additional and technical details are gathered in an Appendix.

%-------------------------------------
\vskip 6pt
{\bf Multifield boost-breaking mixings.---}Using the framework of the EFT of inflationary fluctuations~\cite{Creminelli:2006xe, Cheung:2007st}, we couple the (canonically normalised) Goldstone boson of broken time translations $\pi_c$ to $\Nf$ additional massive scalar fields $\sigma_I$ ($I=1, \ldots, \Nf$). Keeping only time derivatives throughout since spatial gradients do not survive at late times, $k/a\ll H$, and collecting the fields in $\chi^A \equiv (\pi_c, \sigma_I)$, the quadratic Lagrangian we consider is
\begin{equation}
\label{eq: L2 all mixings}
    \L^{(2)} = \frac{1}{2}\K_{AB}\dot\chi^A\dot\chi^B + \rho_I\dot\pi_c\sigma_I + \Omega_{IJ}\dot\sigma_I \sigma_J - \frac{1}{2}M^2_{IJ} \sigma_I\sigma_J \,,
\end{equation}
with $\K_{AB}$ a symmetric matrix with $\K_{00}=1, \K_{0I}=\K_{I0}\equiv \eta_I$, and $\K_{IJ}$ the $\sigma$-sector kinetic matrix, $\rho_I$ an $\Nf$-vector, $\Omega_{IJ}=-\Omega_{JI}$ antisymmetric (the symmetric part is a total time-derivative), and $M^2_{IJ}=M^2_{JI}$ symmetric (here, $\dot{X}\equiv \tfrac{\d X}{\d t}$ with $t$ being cosmic time). We consider all couplings and bare masses to be slow-varying (effectively constant) in time. The Goldstone field is related to the observed curvature perturbation through $\zeta = -H\pi_c /f_\pi^2$, with $f_\pi^4\equiv 2\Mpl^2|\dot H|$ the symmetry-breaking scale (for simplicity, here, we set the speed of sound of the Goldstone to unity, $c_\pi=1$). We use the linear field-redefinition freedom to set $\K_{IJ}=\delta_{IJ}$ (write $\K=LL^T$ and define $\hat\sigma \equiv L^T\sigma$ using positive definiteness, then choose $L=\K^{1/2}$), and use the residual constant $\SO(\Nf)$ rotation freedom to diagonalise the mass matrix $M^2_{IJ}\to \diag(m_I^2)$ (solve the generalised eigenvalue problem $M^2_{IJ}v_a^J = \lambda_a \K_{IJ} v_a^J$ and normalise the eigenvectors so that $v_a^T\K v_b=\delta_{ab}$): this is the classical \emph{normal modes} theorem. This Lagrangian has $\Nf(\Nf+5)/2$ free parameters, and encompasses for example non-linear sigma models and multifield DBI inflation~\cite{Kehagias:1999vr, Langlois:2008wt, Langlois:2008qf, Pinol:2020kvw}.

The quadratic theory must not propagate signals faster than light. Reintroducing the fields' sound speeds $c_\pi$ and $c_I$, \emph{subluminality} for every mode translates to the matrix condition $\N_{AB}\succeq0$, where $\N_{00}=1-c_\pi^2$, $\N_{II}=1-c_I^2$ and $\N_{0I}=\eta_I$. Positive semi-definiteness is equivalent to the single Schur-complement condition:
\begin{equation}
\label{eq: subluminality}
    \sum_{I=1}^{\Nf} \frac{\eta_I^2}{1-c_I^2} \leq 1-c_\pi^2 \,.
\end{equation}
This condition is independent of the \emph{no-ghost bound} $\sum_{I=1}^{\Nf} \eta_I^2<1$ (from $\K\succ0$ alone), and is similar to the ones found in~\cite{Baumann:2011nk, Lee:2016vti}. A non-zero $\eta_I$ only requires $c_\pi$ or $c_{\sigma_I}$ (or both) to depart from unity by a comparable amount $|\eta_I| \lesssim \sqrt{(1-c_\pi^2)(1-c_I^2)}$ at fixed other $\eta_{J\neq I}=0$. In the limiting case where all sound speeds are set exactly to unity, this bound collapses to $\eta_I=0$ for all $I=1, \ldots, \Nf$, which we adopt in what follows.

\begin{figure}[h!] 
    \centering
    \includegraphics[width=0.9\linewidth]{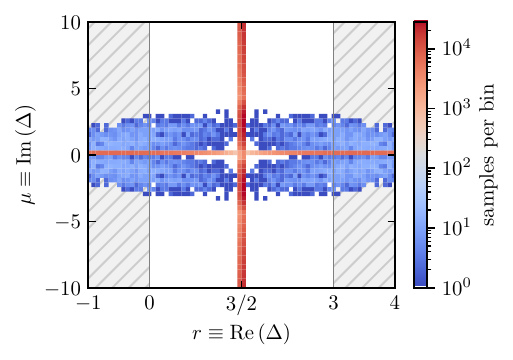}
    \caption{Distribution of the scaling dimensions $\Delta$ in the complex plane by scanning the parameter space for $\Nf=2$. We have drawn $5\times10^5$ samples in the parameter space, with parameters following a uniform distribution with prior $m_I=[-5, 10]H^2$, $\rho_I=[0, 10]H$ and $\omega=[0, 10]H$ ($I=1, 2$).}
    \label{fig: localisation}
\end{figure}

%-------------------------------------
\vskip 6pt
{\bf Scaling dimensions.---}Varying~\eqref{eq: L2 all mixings} with respect to $\pi_c$ and $\sigma_I$ gives the coupled linear equations of motion:
\begin{equation}
\label{eq: multifield EOM}
    \begin{aligned}
    &\ddot\pi_c + 3H\dot\pi_c + \rho_I(3H\sigma_I + \dot\sigma_I) = 0 \,, \\
    &\ddot\sigma_I + 3H\dot\sigma_I + 2(\Omega\dot\sigma)_I + 3H(\Omega\sigma)_I + m_I^2\sigma_I = \rho_I\dot\pi_c \,,
    \end{aligned}
\end{equation}
with $(\Omega\dot\sigma)_I \equiv \Omega_{IJ}\dot\sigma_J$. Substituting the power-law ansatz $\chi^A = \bar\chi^A(-\tau)^\Delta$ in conformal time $\tau$ ($\d\tau\equiv \d t/a(t)$) into~\eqref{eq: multifield EOM} turns the system into the algebraic condition $\M(\Delta)\bar\chi = 0$, with
\begin{align}
\label{eq: scaling-dimension matrix}
    \M_{00} &= \Delta(\Delta-3) \,, \quad \M_{0I} = \frac{\rho_I}H(3-\Delta) \,, \quad \M_{I0} = \frac{\rho_I}H\Delta \,, \nonumber \\
    \M_{IJ} &= \Big[\Delta(\Delta-3)+\frac{m_I^2}{H^2}\Big]\delta_{IJ}+\frac{\Omega_{IJ}}H(3-2\Delta)\,.
\end{align}
A non-trivial solution ($\bar\chi\neq0$) requires $\det \M(\Delta)=0$: the late-time scaling dimensions $\Delta$ are the roots of this degree-$2(\Nf+1)$ characteristic polynomial. The roots $\Delta=0,3$---corresponding to the constant and decaying $\pi_c$-modes---are naturally exact for any $\Nf$ and any values of masses and mixing couplings.

Mixings can drive the effective masses of additional fields tachyonic, pushing $\Delta$ below zero. This signals a super-horizon instability, as the corresponding mode continues to grow outside the horizon. Requiring \emph{effective mass positivity} translates to the algebraic requirement:
\begin{equation}
\label{eq: meff positivity}
    \min_a\,{\Re}\big(\Delta_a\big) \geq 0 \,,
\end{equation}
over all $2(\Nf+1)$ roots of $\det\M(\Delta)=0$ (with $\Delta=0, 3$ automatically saturating it from the $\pi_c$ sector). Notice that $M_{IJ}^2\succeq0$ is neither necessary nor sufficient: mixings can drive an individually positive mass matrix unstable, and conversely well-chosen mixings can stabilise directions that would be alone tachyonic.

\begin{figure}[h!]
    %\hspace*{-1.5cm}
    \centering
    \includegraphics[width=\linewidth]{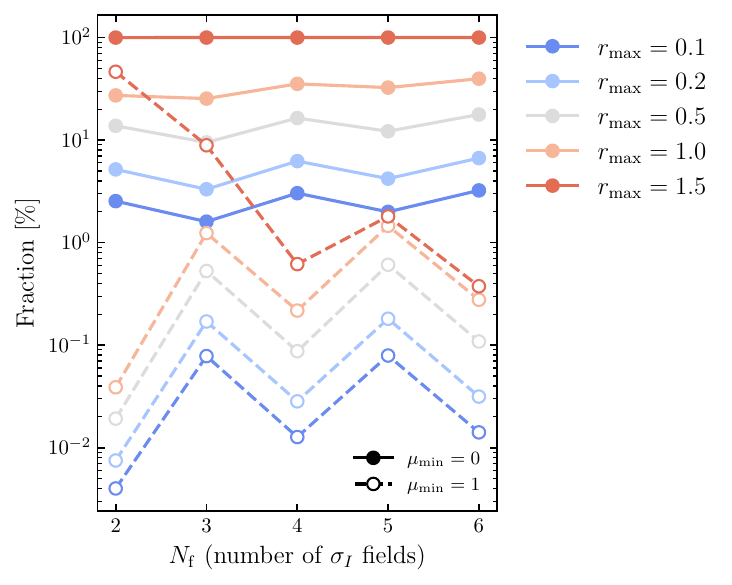}
   \caption{Fraction of natural-scale parameter space with $\Delta=r+i\mu$, such that $r\leq r_{\max}$ and $\mu_{\min}\leq\mu$, as a function of the number $\Nf$ of additional fields. We have uniformly drawn $10^6$ samples with prior $m_I^2=[-5, 10]H^2$ and $\rho_I, \Omega_{IJ}=[-10, 10]H$ ($I, J=1, \ldots, \Nf$). The oscillating ($\mu_{\rm min}=1$) fractions alternate sharply with the parity of $\Nf$, tracking whether $\Omega_{IJ}$ is forced singular (odd $\Nf$) or not (even $\Nf$). We have removed the ever-present $\Delta=0, 3$ massless modes.}
  \label{fig: multifield fractions}
\end{figure}

%-------------------------------------
\vskip 6pt
{\bf Fine-tuned signals for $\Nf=\O(1)$.---} For $\Nf=2$, the quadratic theory~\eqref{eq: L2 all mixings} has 5 parameters: $m_{1, 2}, \rho_{1, 2}$ and $\omega$ (recall $\eta_I=0$ throughout). The equation $\det\M(\Delta)=0$ can be solved explicitly:
\begin{equation}
\label{eq: scaling dimensions N=2}
    \Delta = \frac{3}{2}\pm\sqrt{\frac{9}{4}-\frac{B}{2} \pm \frac{\sqrt{B^2-4C}}{2}} \,,
\end{equation}
where $B = [\sum_I (m_I^2+\rho_I^2) + 4\omega^2]/H^2$ and $C = [m_1^2m_2^2 + \rho_1^2m_2^2 + \rho_2^2m_1^2 + 9H^2\omega^2]/H^2$. For $\Nf\geq3$, $\det\M(\Delta)=0$ has no closed-form solution and we resort to numerics. We randomly scan the parameter space by uniformly sampling all parameters, allowing the bare masses to be tachyonic.  
Fig.~\ref{fig: localisation} shows the histogram of scaling dimensions $\Delta\equiv r+i\mu$ in the complex plane $(r, \mu)$. The gray regions signal an effective tachyonic mode in the theory, which therefore is considered unhealthy. Remarkably, we observe a phenomenon of \emph{localisation}: even in the presence of boost-breaking mixings, the scaling dimensions remain clustered around the unitary irreducible representations of the dS group (the shadow symmetry manifests as a $\mathbb{Z}_2$ symmetry around the axis $r=3/2$ and $\mu=0$). Specifically, the sampled scaling dimensions agglomerate around either the complementary series or the principal series. This effect is also observed for $\Nf\geq3$. The scaling dimensions exhibit a finite spread around the complementary series, with a moderate width of $\delta\mu\approx 3$, while localisation around the principal series remains strict, with $\delta r=0$. Eventually, we observe that the region $0<r\equiv\Re(\Delta)\ll1$ is populated, with the possibility of having $\mu\equiv\Im(\Delta)\neq0$ (here up to $\mu\approx3$ for rare events). This region produces \emph{stable} ($\Re(\Delta)>0$ for the dominant root), \emph{enhanced} ($\Re(\Delta)\lesssim1$), and possibly \emph{oscillating} ($\Im(\Delta)\neq0$) cosmological collider signals in the squeezed bispectrum. 

\begin{figure}[h!]
    %\hspace*{-1.5cm}
    \centering
    \includegraphics[width=\linewidth]{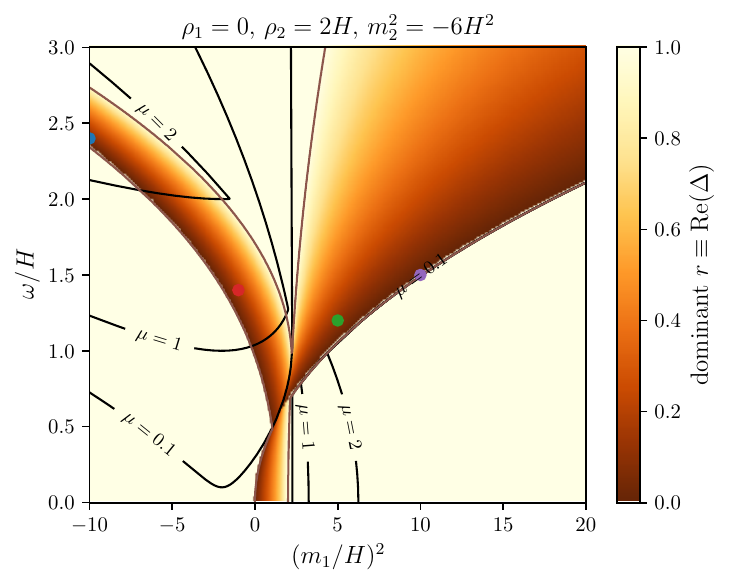}
   \caption{Phase diagram in the $((m_1/H)^2, \omega/H)$ plane at fixed $\rho_1=0, \rho_2=2H, m_2^2=-6H^2$ ($\Nf=2$). The colour map shows the dominant real part of the scaling dimensions $r\equiv\Re(\Delta)$ (above the capped $r=1$, $r$ saturates at $3/2$). The black lines label $\mu\equiv\Im(\Delta)$ contours. Colour dots mark points for which we compute the full bispectrum shape, shown in Fig.~\ref{fig: full shapes}.}
  \label{fig: phase diagram}
\end{figure}

In Fig.~\ref{fig: multifield fractions}, we quantify how much of the parameter space can accommodate for such signals: $\Delta=r+i\mu$ for $r\leq r_{\rm max}$, and with $\mu\geq\mu_{\rm min}$. For $\Nf=\O(1)$, a generically enhanced signal ($r\leq1$) represents up to $\approx 30\%$ of the parameter space, while requiring the signal to approach the local shape ($r\ll1$) makes the fraction drop to $\O(1\%)$. Selecting oscillating signals ($\mu_{\rm min}\geq1$) makes the fraction further drop below $\O(0.1\%)$: such signals are therefore highly \emph{fine-tuned}. The dashed curves (i.e.~requiring genuine oscillations, $\mu\geq1$) alternate sharply with the parity of $\Nf$, with odd $\Nf$ generating oscillating collider signals systematically more often than even $\Nf$ by up to an order of magnitude at matched thresholds. This can be traced to a simple mechanism: an $\Nf\times\Nf$ antisymmetric matrix is exactly singular whenever $\Nf$ is odd (since $\det\Omega = (-1)^{\Nf}\det\Omega$), so $\Omega_{IJ}$ is forced to have a vanishing eigenvalue, and hence a genuine null direction in field space, for every odd $\Nf$. Moreover, genuinely resolvable heavy-field oscillations, with $r\geq3/2$ and $\mu\geq1$ (deep in the principal series), are also fine-tuned, occupying only $\sim1\%$ of the parameter space for $\Nf\geq4$ (as a sanity check, we verify that the broader region $r\geq3/2$, $\mu\geq0$ encompasses the full parameter space).

\begin{figure*}[t]
    \centering
    \includegraphics[width=0.42\textwidth]{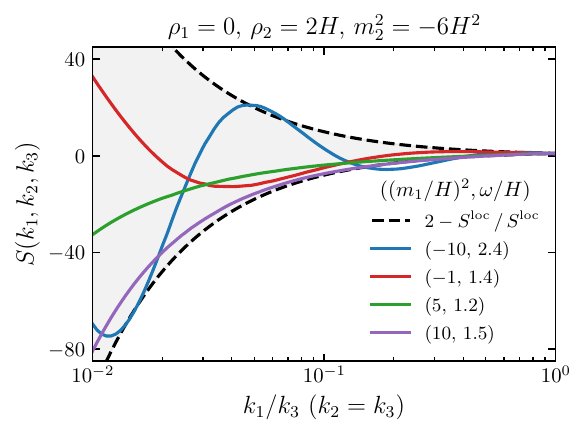}
    %\hfill
    \hspace*{1.5cm}
    \includegraphics[width=0.38\textwidth]{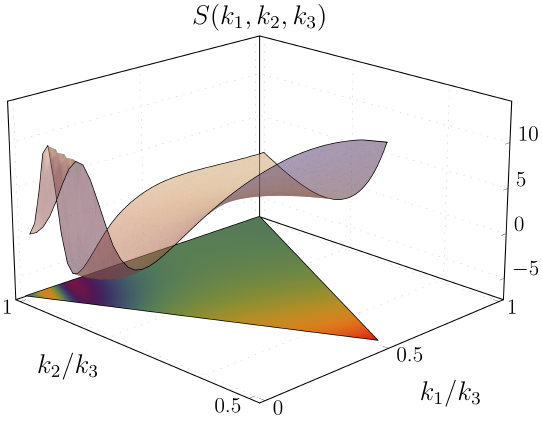}
    \caption{{\it Left panel}: Squeezed limit of the bispectrum shape generated by all cubic interactions fixed by the non-linearly realised symmetry in the isosceles triangle configuration at fixed $\rho_1=0, \rho=2H, m_2^2=-6H^2$, for different values of $(m_2/H)^2$ and $\omega/H$ represented in Fig.~\ref{fig: phase diagram} ($\Nf=2$), and normalised to unity in the equilateral limit. The shaded region corresponds to the limiting local shapes, marking the unitarity boundary ($\Re(\Delta)=0$). {\it Right panel}: Full bispectrum shape $S(k_1, k_2, k_3)$ in all kinematic configurations for $m_1^2=-10H^2, m_2^2=-6H^2, \rho_1=0, \rho_2=2H$ and $\omega=2.4H$ (same as blue dot in the squeezed limit). The squeezed limit has been truncated to $k_3/k_1=0.03$, but continues growing to approach the local shape.}
    \label{fig: full shapes}
\end{figure*}

We localise these signals in parameter space in Fig.~\ref{fig: phase diagram}, for $\Nf=2$. For this choice of fixed parameters, we observe two branches for which $r<0.5$, giving enhanced bispectrum squeezed limits. The left branch exhibits oscillations, as shown by the $\mu$ levels, whereas the right branch represents pure power-law signals. Enhanced signals are more present for tachyonic bare masses, and the frequency of oscillations increases as the twist $\omega$ increases. For selected parameters, we show in Fig.~\ref{fig: full shapes} the corresponding dimensionless shape function $S(k_1, k_2, k_3)\equiv (k_1k_2k_3)^2 B_\zeta/[(2\pi)^4\P_\zeta^2]$ (where $\P_\zeta=A_s\sim 10^{-4}$ is the primordial power spectrum amplitude, and $B_\zeta$ is the primordial bispectrum) in all kinematic configurations, numerically computed with {\sf CosmoFlow}~\cite{Werth:2023pfl, Pinol:2023oux, Werth:2024aui}. We have included all cubic interactions fixed by the non-linearly realised symmetry: $\sigma_I(\partial_i\pi)^2/a^2$ fixed by $\rho_I$ and $\sigma_I(\partial_i \pi)(\partial_i \sigma_J)/a^2$ fixed by $\omega$ (recall that we set $\eta_I=0$). The Goldstone boson self-interactions are fine-tuned to zero. These shapes exhibit an enhanced (almost local) squeezed limit, with oscillations in the case of both bare masses being tachyonic. Since $\Delta_++\Delta_-=3$ by Vieta's relation, whenever $r_{\text{dom}}\equiv \min_a\,{\Re}\big(\Delta_a\big)\ll1$, the competing mode sits at $\approx 3-r_{\text{dom}}\approx3$, automatically far away, so that the clean signal is not spoiled by a comparably dominant mode. Looking at the full shape, we notice an enhanced folded limit, which reflects the transient tachyonic instability of additional fields (here experienced by both $\sigma_I$ fields since their bare masses are negative)~\cite{Fumagalli:2019noh, Aoki:2026qea}. 

%-------------------------------------
\vskip 6pt
{\bf Cosmological colliders at large-$\Nf$.---}At $\Nf=\O(1)$, an enhanced squeezed limit is a fine-tuned corner of parameter space. We now ask what happens as $\Nf$ grows, by drawing the bare mass spectrum $(m_I/H)^2$, the twist matrix $\Omega_{IJ}/H$, and the linear mixing $\rho_I/H$ independently from natural ensembles and tracking the dominant root $r_{\rm dom}$ (excluding the exact $\Delta=0,3$ roots) as function of $\Nf$.

\begin{figure}[h!]
    %\hspace*{-1.5cm}
    \centering
    \includegraphics[width=0.9\linewidth]{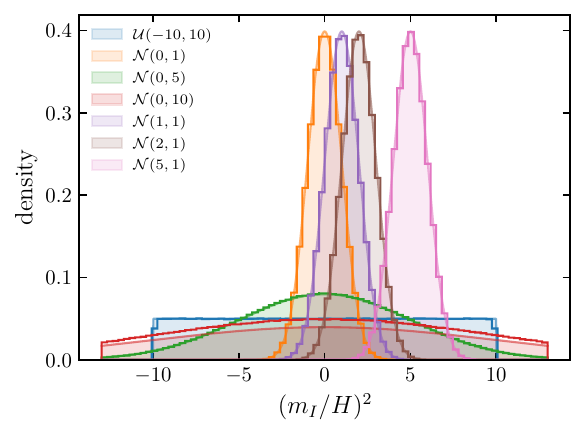}
   \caption{The seven mass priors $(m_I/H)^2$ compared in the large-$\Nf$ scan: a uniform prior, three mean-zero normals of increasing width, and three normals displaced from zero (``localised'').}
  \label{fig: mass distributions}
\end{figure}

For the mass spectrum, we compare a uniform prior $(m_I/H)^2\sim\U(-10, 10)$, three mean-zero normals $\N(0, \sigma)$ with $\sigma=1, 5, 10$, and three ``localised'' normals $\N(c, 1)$ offset from zero by $c=1, 2, 5$ (Fig.~\ref{fig: mass distributions}). For $\Omega_{IJ}/H$, we build a real antisymmetric matrix with entries weighted by an exponential correlation length $\ell$, $\Omega_{IJ}/H\propto e^{-|I-J|/\ell}$, interpolating from strongly banded ($\ell=1$, only near-neighbour fields $\sigma_I$ mix) to fully dense ($\ell\to\infty$), alongside the limiting case $\Omega_{IJ}=0$. The linear mixings $\rho_I/H$ are drawn from the same family of distributions used for the masses. 

\begin{figure}[h!]
    %\hspace*{-1.5cm}
    \centering
    \includegraphics[width=\linewidth]{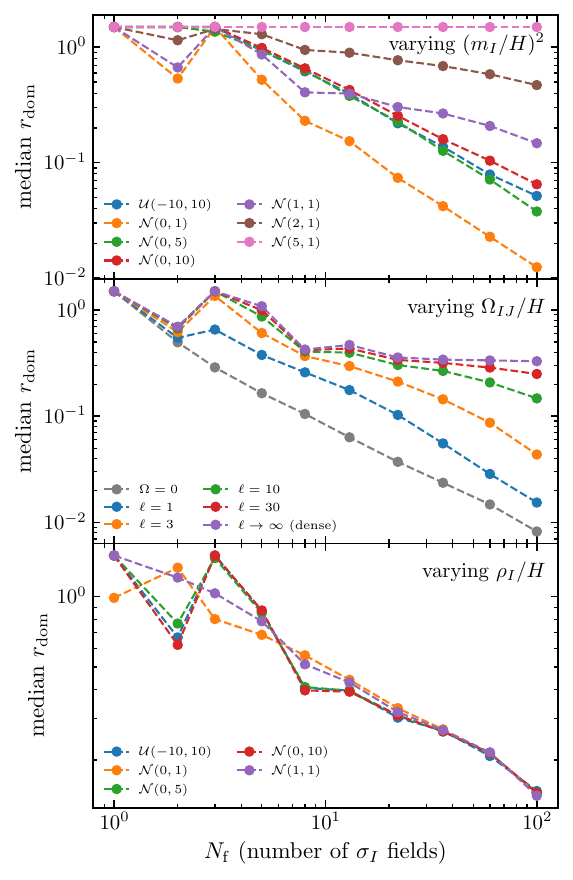}
    \caption{Median dominant root $r_{\mathrm{dom}}\equiv\min_a\Re(\Delta_a)>0$ as a function of $\Nf$, varying the mass distribution (\textit{top}), $\Omega_{IJ}$ (\textit{middle}), and the $\rho_I$ distribution (\textit{bottom}), at fixed conventions for the other two, using $10^6$ samples per point. All three collapse towards small $r_{\rm dom}$ as $\Nf\to100$ for generic priors (only mass spectra offset from the mixing scale by several widths fail to enhance cosmological collider signals).}
  \label{fig: large-Nf}
\end{figure}

Fig.~\ref{fig: large-Nf} is the main result. For essentially every distribution tested, $\text{median}(r_{\rm dom})$ falls as a clean power law in $\Nf$. The precise exponent depends on where the mass spectrum sits relative to the mixing scale and on how densely $\Omega_{IJ}$ couples the fields, but strikingly not on the statistical law used to draw the mixings $\rho_I$. The only priors that do not generate enhanced cosmological collider signals are localised heavy bare mass spectra displaced from zero by several mixing-widths (e.g.~$\N(5, 1)$ stays pinned at $r=3/2$ for every $\Nf$). In this case, the linear mixings pulling the dominant scaling dimension towards the unitarity boundary $\Delta=0$ is a rare event.

These results are a direct consequence of extreme value statistics. By construction, $r_{\mathrm{dom}} = \min_a\Re(\Delta_a)$ is the minimum of $\O(\Nf)$ roots of $\det\M(\Delta)=0$. Whenever the pooled non-trivial root density behaves as $g_{\Nf}(r)\sim g_0 r^\kappa$ near $r=0^+$, for some finite $\kappa>-1$, the Fisher--Tippett--Gnedenko theorem fixes the decay of the minimum of $\Nf$ such that: $\text{median}(r_{\rm dom})\sim \Nf^{-1/(1+\kappa)}\to0$, with the exponent set purely by the \emph{local} exponent $\kappa$ of the density at the unitarity boundary $r=0$, not by the distribution's global shape. We verify this directly by fitting $\kappa$ from the pooled root density at $\Nf=100$, which predicts the exponent measured across the full scan to $\sim 12\%$. The one exception is when the bare mass distribution is localised around heavy masses, and is precisely the case where $\kappa\to\infty$, i.e.~a genuine gap opens in the density at $r=0$.

Physically, the $\Nf=\O(1)$ picture inverts. Any UV completion that generates a large number of fields generically produces an enhanced, almost-local bispectrum shape, without tuning of the masses or mixings. Oscillations, by contrast, remain just as rare as for $\Nf=\O(1)$. We do not expect $\eta$-type mixings (i.e.~$\eta_I\,\dot\pi_c\sigma_I$) to significantly change this picture since subluminality constraints these couplings to be parametrically small, even when reduced sound speeds are turned on, see Eq.~\eqref{eq: subluminality}.

%-------------------------------------
\vskip 6pt
{\bf Discussion.---}Mapping observable signatures back to the underlying theory is a major challenge. Cosmological correlators are shaped by many degeneracies and competing effects, which obscure the field masses, couplings, and interactions we ultimately want to extract. Cosmological collider signals are no exception. In this \emph{Letter}, we have shown that the standard distinction between heavy- and light-field signatures can be radically altered by the number of fields active during inflation and by their linear mixings. In particular, for large $\Nf=\O(100)$, the median real part of the dominant scaling dimension exhibits a clean power-law decay with $\Nf$, with the exponent determined by the underlying mass and mixing distributions. This reveals a universal tendency for mixings to drive the scaling dimensions toward the unitarity boundary, making almost-local cosmological collider signals a generic prediction. This has direct implications for searches in the data: almost-local collider signals are free from equilateral contamination from inflaton self-interactions, and can be efficiently probed through scale-dependent bias, making constraints on $\fnl^{\rm loc}$ a natural discriminator of new physics involving many species during inflation.\looseness=-1 

The robustness of this picture is, of course, tied to its assumptions. It would therefore be interesting to narrow down the mass and mixing priors using positivity bounds and naturalness considerations, and identify concrete inflationary backgrounds capable of generating the mixings considered here. 

%-------------------------------------
\vskip 6pt
\begin{acknowledgments}
{\it Acknowledgments.---} We used Claude (Opus 5.5, Anthropic, 2026) as a coding assistant for parts of the numerical implementation. DW is funded by the Max Planck-IAS-NTU Center for Particle Physics, Cosmology and Geometry, by the Deutsche Forschungsgemeinschaft (DFG, German Research Foundation) under Germany’s Excellence Strategy---EXC-2094/2-390783311, and by the European Union (ERC, UNIVERSE+, 101118787). {\small Views and opinions expressed are however those of the author(s) only and do not necessarily reflect those of the European Union or the European Research Council Executive Agency. Neither the European Union nor the granting authority can be held responsible for them.}
\end{acknowledgments}

%-------------------------------------
\bibliographystyle{utphys}
\bibliography{references}

%-------------------------------------
\appendix
\section{Appendix}
\label{sec: app}

In this appendix, we follow the notation of the main text.  We first consider a two-field model as a warm-up exercise. Then, we construct the most general linear mixings from the unitary gauge, and determine cubic operators that are fixed by the non-linearly realised symmetry. Eventually, we discuss the numerical implementation for solving for the late-time scaling dimensions, and provide details about extreme value statistics in random matrix~theory.

%-------------------------------------
\vskip 6pt
{\bf Collider signals from two-field mixings.---}Beyond reproducing known and standard results, this elementary example serves as an illustration for the general strategy adopted throughout this work. Let us consider the standard linear ``turn'' mixing, for which the quadratic Lagrangian reads:
\begin{equation}
    \L^{(2)} = -\frac{1}{2}(\partial_\mu \pi_c)^2 - \frac{1}{2}\left[(\partial_\mu \sigma)^2 + m^2 \sigma^2\right] + \rho \, \dot\pi_c\sigma \,.
\end{equation}
Here, we define the Lagrangian density by stripping off the factor $\sqrt{-g}=a^3(t)$: $S=\int \d t \d^3\x \, a^3(t) \, \L$, and use mostly-plus metric signature, so that $-\frac12(\partial_\mu X)^2=\frac12\dot X^2-\frac{1}{2a^2}(\partial_iX)^2$ for either field. The coupled linear equations of motion are given by
\begin{equation}
\label{eq: pi-sigma model EOMs}
    \left\{
    \begin{aligned}
        &\ddot\pi_c + 3H\dot\pi_c + \frac{k^2}{a^2}\pi_c = -\rho\left(3H\sigma+\dot\sigma\right) \,, \\
        &\ddot\sigma + 3H\dot\sigma + \left(\frac{k^2}{a^2}+m^2\right)\sigma = \rho\dot\pi_c \,.
    \end{aligned}
    \right.
\end{equation}
At late-times, equivalently on super-horizon scales at the mode level, $k/a\ll H$, the gradient terms become negligible compared to the friction, mixing and mass terms. This is precisely the regime that controls the squeezed limit of the bispectrum (and more generally soft limits of high-order point correlators), since $k_3\ll k_1, k_2$ means that the mode $k_3$ exits the horizon parametrically earlier and spends more time deep in this regime. Substituting the power-law ansatz $\pi_c = \bar\pi_c \, \tau^\Delta$ and $\sigma = \bar\sigma \, \tau^\Delta$ into the equations of motion~\eqref{eq: pi-sigma model EOMs} yields $\M(\Delta)
    \begin{pmatrix}
        \bar\pi_c \\ 
        \bar\sigma
    \end{pmatrix}=0$, with
\begin{equation}
    \M(\Delta)=
    \begin{pmatrix}
        \Delta(\Delta-3) & \dfrac{\rho}{H}(3-\Delta)\\[6pt]
        \dfrac{\rho}{H}\Delta & \Delta(\Delta-3)+\dfrac{m^2}{H^2}
    \end{pmatrix} \,.
\end{equation}
A non-trivial solution $(\bar\pi_c, \bar\sigma) \neq (0, 0)$ requires $\det\M(\Delta)=0$. Solving the quartic equation gives:
\begin{equation}
    \Delta_-^\pi=0 \,,\quad \Delta_+^\pi=3 \,,\quad
    \Delta_\pm^\sigma = \frac32\pm i\sqrt{\frac{\meff^2}{H^2}-\frac94} \,,
\end{equation}
where $\meff^2 \equiv m^2+\rho^2$ is the effective mass on super-horizon scales~\cite{Castillo:2013sfa, An:2017hlx, Iyer:2017qzw, Pinol:2023oux}. As a trivial consistency check, notice that the scaling-dimension matrix becomes diagonal when $\rho=0$. The factor $\Delta(\Delta-3)$ is completely independent of the mixing $\rho$: the constant ($\Delta=0$) and decaying ($\Delta=3$) modes of the curvature perturbation survive as {\it exact} solutions for any mixing strength.

\begin{table*}[t!]
\begin{center}
    \renewcommand{\arraystretch}{1.2}
    \begin{threeparttable}
    \begin{tabular}{@{}lll@{}}
    \hline\hline
    Matrix & Unitary-gauge operator & Description \\
    \hline
    $\K_{00}$ & $\tfrac12 \Mpl^2R+\Mpl^2\dot{H} g^{00} - \Mpl^2(3H^2+\dot{H})$ & standard single-clock operator \\
    $\K_{00}$ & $M_2(t)^4\,(\delta g^{00})^2$ & single-clock operator $(c_\pi\neq1)$ \\
    $\K_{IJ}$ (part) & $-e_1^{IJ}(t)\,g^{\mu\nu}\partial_\mu\sigma_I\partial_\nu\sigma_J$ & Lorentz-invariant operator \\
    $\K_{IJ}$ (part) & $e_2^{IJ}(t)\,(g^{0\mu}\partial_\mu\sigma_I)(g^{0\mu}\partial_\mu\sigma_J)$ & boost-breaking piece ($c_I\neq1$) \\
    $\K_{0I}$ ($\eta_I$) & $\tilde M_1^{2I}(t)\,\delta g^{00}\,(g^{0\mu}\partial_\mu\sigma_I)$ & shift-symmetric $\eta_I$ operator \\
    $\rho_I$ & $-\tilde M^{3I}(t)\,\delta g^{00}\,\sigma_I$ & requires $\sigma_I$ \emph{not} shift-symmetric \\
    $\Omega_{IJ}$ & $\Omega_{IJ}(t)\,(g^{0\mu}\partial_{\mu}\sigma_I)\sigma_J$ & requires $\sigma_I$ \emph{not} shift-symmetric \\
    $M^2_{IJ}$ & $-\tfrac12 M^2_{IJ}(t)\,\sigma_I\sigma_J$ & Lorentz-invariant operator \\
    \hline\hline
    \end{tabular}
    \caption{Classification of unitary-gauge operators that generate the mixings in the quadratic Lagrangian~\eqref{eq: L2 all mixings} up to two (time) derivatives.}
    \label{tab: unitary-gauge operators}
    \end{threeparttable}
\end{center}
\end{table*}

\vskip 6pt
{\it Adding mixings.---}We now include the remaining independent Lorentz-breaking mixing between $\pi_c$ and $\sigma$ involving at most one time derivative: $\L^{(2)} \supset \eta \, \dot\pi_c\dot\sigma$. Writing the kinetic term as $\tfrac{1}{2}\dot{\chi}^T \K \dot{\chi}$ where $\chi\equiv (\pi_c, \sigma)$ and $\K = \begin{pmatrix} 1&\eta\\\eta&1 \end{pmatrix}$ with eigenvalues $1\pm\eta$, positivity (no ghost) requires $|\eta|<1$. Since the mixing $\eta \, \dot\pi_c\dot\sigma$ only multiplies $\dot\pi_c$, we can complete the square and perform the following local field redefinition to remove it: $\pi_c \to \pi_c + \eta \, \sigma$. If the field $\sigma$ is effectively sufficiently heavy, it decays enough on super-horizon scales so that $\pi_c$ is still linearly related to the curvature perturbation $\zeta$. We obtain $\frac{1}{2}\dot\pi_c^2 + \eta \, \dot\pi_c\sigma \to \frac{1}{2}\dot\pi_c^2 - \frac{1}{2}\eta^2\dot\sigma^2$. The cross-term $\dot\pi_c\dot\sigma$ cancels exactly, and $\pi_c$ is left with a canonical kinetic term. Notice that this field redefinition also generates mixed gradient terms, of the form $\delta^{ij}(\partial_i \pi_c)(\partial_j \sigma)/a^2$, but we omit them since they do not contribute at late times. The remaining mixing becomes $\rho \, \dot\pi_c\sigma \to \rho \, \dot\pi_c\sigma - \rho\eta \, \sigma\dot\sigma$. Since $\sigma\dot\sigma = \tfrac{1}{2}\tfrac{\d\sigma^2}{\d t}$ is a total derivative, integrating it by parts against the $a^3(t)$ measure ($\int a^3 \dot{f}\d t = -\int 3Ha^3 f \d t+\text{bdy}$) turns it into an ordinary mass term, $\tfrac{3}{2}H\rho\eta\sigma^2$. Canonically normalising $\tilde\sigma \equiv \sqrt{1-\eta^2} \, \sigma$, the late-time Lagrangian is exactly the old canonical form, with shifted parameters
\begin{equation}
    \tilde{m}^2 = \frac{m^2-3H\rho\eta}{1-\eta^2}\,, \qquad \tilde{\rho} = \frac{\rho}{\sqrt{1-\eta^2}} \,.
\end{equation}
The scaling dimensions are therefore given by $\Delta^\pi_\pm=0, 3$ and $\Delta_\pm^\sigma = \frac32 \pm i\sqrt{\frac{\meff^2(\eta)}{H^2} - \frac94}$ with
\begin{equation}
    \meff^2(\eta) \equiv \tilde{m}^2 + \tilde\rho^2 = \frac{m^2+\rho^2-3H\rho\eta}{1-\eta^2} \,.
\end{equation}
These scaling dimensions can also be recovered from the equations of motion at late-times, without performing any field redefinition. The new effective mass $\meff(\eta)$ brings new phenomenology to the linear sector. The cross-term $-3H\rho\eta$ is a genuine interference effect between the two mixings, and can have either sign. Unlike the $\rho$-only case, where the mixing could only push the physical $\sigma$ mode towards the principal series, a suitable $\eta$ can now decrease $\meff^2(\eta)$. For example at $m^2=0$, we have $\meff^2(\eta) = \rho(\rho-3H\eta)/(1-\eta^2)$ vanishes at $\eta=\rho/(3H)$ and even turns negative beyond it, so long as $\rho<3H$ keeps this crossing inside the healthy range $|\eta|<1$. A negative $\meff^2$ sends $\Delta_-^\sigma$ below zero: a genuinely growing super-horizon mode. Eventually, notice that adding gradient mixing terms like $\L^{(2)} \supset \bar\rho \, \delta^{ij}(\partial_i \pi_c)(\partial_j \sigma)/a^2$ do not survive at late times, i.e.~in the limit $k\to0$. Even though such terms matter physically (for example for the sub-horizon evolution of the corresponding mode functions, and therefore for the overall power spectrum normalisation, and the non-Gaussian size), we discard them.

Generalising this to multiple additional fields proceeds along essentially the same lines. The main complication beyond this elementary example is the rapid proliferation of mixing parameters, which renders the determination of the scaling dimensions a numerical problem.

%-------------------------------------
\vskip 6pt
{\bf Unitary-gauge operators.---}We review how the mixings in Eq.~\eqref{eq: L2 all mixings} arise from the EFT of inflationary fluctuations perspective. Unitary gauge fixes a scalar clock $\tau(x)$ such that $\tau=t$, so that the Goldstone boson of broken time translations is eaten by the metric. Reintroducing the Goldstone boson to restore full diffeomorphism invariance is done by the St\"uckelberg trick. Any unitary-gauge operator built from $g^{\mu\nu}\partial_\mu\tau\partial_\nu\tau=g^{00}$ (invariant under spatial diffeomorphism) St\"uckelberg-substitutes simply by this replacement $g^{00} \to g^{\mu\nu} \partial_\mu(t+\pi) \partial_\nu(t+\pi) = g^{00} + 2g^{0\mu}\partial_\mu\pi + g^{\mu\nu}\partial_\mu\pi\partial_\nu\pi$, which at the background level ($g^{00}=-1, g^{0i}=0, g^{ij}=\delta^{ij}/a^2$) gives
\begin{equation}
    \delta g^{00} \equiv g^{00}+1 \to -2\dot\pi-\dot\pi^2+\frac{(\partial_i\pi)^2}{a^2} \,.
\end{equation}
Each building term in Eq.~\eqref{eq: L2 all mixings} traces to a specific unitary-gauge operator, built from $\delta g^{00}\equiv g^{00}+1$ and $g^{0\mu}\partial_\mu(\cdot)$, or requiring no dressing at all for combinations of $\sigma_I$ ($I=1, \ldots, \Nf$) that transform as scalars under full diffeomorphisms, or $g^{0\mu}\partial_\mu\sigma_I$ and $g^{\mu\nu}(\partial_\mu\sigma_I)(\partial_\nu\sigma_J)$. Following a (time) derivative expansion up to second-order only, unitary-gauge operators contributing to the quadratic Lagrangian for fluctuations are listed in Tab.~\ref{tab: unitary-gauge operators}. The standard single-clock operator is entirely fixed by requiring that tadpoles vanish so that the action starts quadratic in the fluctuations. Fine-tuning the speed of sound of $\pi$ to be unity amounts to setting $M_2=0$. Without loss of generality, we can always set $e_1^{IJ}=\delta^{IJ}$ and $e_2^{IJ}$ to be diagonal. Further setting $e_2^{IJ}=0$ fixes the speed of sounds of $\sigma_I$ to unity. Shift-symmetric operators in $\sigma_I$ were all classified in~\cite{Senatore:2010wk}, and the discussion has been extended to additional heavy fields in~\cite{Noumi:2012vr} (see also~\cite{Pinol:2024arz}). Importantly, notice that this construction does not allow gradient mixings of the form $\delta^{ij}(\partial_i \pi)(\partial_j \sigma_I)$, since $\pi$ non-linearly realises time diffeomorphisms.

\begin{table*}[t!]
\begin{center}
    \renewcommand{\arraystretch}{1.2}
    \begin{threeparttable}
    \begin{tabular}{@{}ll@{}}
    \hline\hline
    Unitary-gauge operator & Reintroducing the Goldstone boson \\
    \hline
    $(\delta g^{00})^2$ & $4\dot\pi+4\dot\pi^2-\textcolor{pyblue}{4\dot\pi(\partial_i\pi)^2/a^2}+\cdots$ \\
    $\delta g^{00} \, \sigma_I$ & $-2\dot\pi\sigma_I-\dot\pi^2\sigma_I+\textcolor{pyblue}{(\partial_i\pi)^2/a^2\sigma_I}$ \\
    $\delta g^{00} \,(g^{0\mu}\partial_\mu\sigma_I)$ & $2\dot\pi\dot\sigma_I+3\dot\pi^2\dot\sigma_I-\textcolor{pyblue}{\dot\sigma_I(\partial_i\pi)^2/a^2}-\textcolor{pyblue}{2\dot\pi(\partial_i\pi)(\partial_i\sigma_I)/a^2}+\cdots$ \\
    $(g^{0\mu}\partial_\mu\sigma_I)(g^{0\mu}\partial_\mu\sigma_J)$ & $\dot\sigma_I\dot\sigma_J+\dot\pi\dot\sigma_I\dot\sigma_J-\textcolor{pyblue}{\dot\sigma_I(\partial_i\pi)(\partial_i \sigma_J)/a^2}+ (I\leftrightarrow J)+\cdots$ \\
    $(g^{0\mu}\partial_\mu\sigma_I)\sigma_J$ & $-\dot\sigma_I\sigma_J-\dot\pi\dot\sigma_I\sigma_J+\textcolor{pyblue}{\sigma_J(\partial_i\pi)(\partial_i\sigma_I)/a^2} + (I\leftrightarrow J)$ \\
    \hline
    $(\delta g^{00})^2\sigma_I$ & $4\dot\pi^2\sigma_I + \cdots$ \\
    $(\delta g^{00})^2 (g^{0\mu}\partial_\mu\sigma_I)$ & $-4\dot\pi^2\dot\sigma_I+\cdots$ \\
    $\delta g^{00} \, g^{\mu\nu}(\partial_\mu \sigma_I)(\partial_\nu \sigma_J)$ & $-2\dot\pi\left[-\dot\sigma_I\dot\sigma_J + (\partial_i \sigma_I)(\partial_i \sigma_J)\right]+\cdots$ \\
    $\delta g^{00} \sigma_I \sigma_J$ & $-2\dot\pi\sigma_I\sigma_J+\cdots$ \\
    $\delta g^{00} (g^{0\mu}\partial_\mu\sigma_I)(g^{0\mu}\partial_\mu\sigma_J)$ & $-2\dot\pi\dot\sigma_I\dot\sigma_J+\cdots$ \\
    $\delta g^{00} \sigma_I (g^{0\mu}\partial_\mu\sigma_J)$ & $2\dot\pi\sigma_I\dot\sigma_J+\cdots$ \\
    \hline
    $(\delta g^{00})^3$ & $-8\dot\pi^3+\cdots$ \\
    $\sigma_I\sigma_J\sigma_K$ & $\sigma_I\sigma_J\sigma_K$ \\
    $\sigma_I\sigma_J(g^{0\mu}\partial_\mu\sigma_K)$ & $-\sigma_I\sigma_J\dot\sigma_K+\cdots$ \\
    $\sigma_I(g^{0\mu}\partial_\mu\sigma_J)(g^{0\mu}\partial_\mu\sigma_K)$ & $\sigma_I\dot\sigma_J\dot\sigma_K+\cdots$ \\
    $\sigma_I \, g^{\mu\nu}(\partial_\mu\sigma_J)(\partial_\nu\sigma_K)$ & $\sigma_I\left[-\dot\sigma_J\dot\sigma_K + (\partial_i\sigma_J)(\partial_i\sigma_K)/a^2\right]$ \\
    \hline\hline
    \end{tabular}
    \caption{Unitary-gauge operators generating cubic interactions for the fluctuations after performing the St\"uckelberg trick. The first block corresponds to operators also generating the quadratic Lagrangian, the second block generates mixed cubic operators, and the last block generates cubic self-interactions. Operators fixed by the non-linearly realised symmetry are highlighted in \textcolor{pyblue}{blue}.}
    \label{tab: unitary-gauge operators for L3}
    \end{threeparttable}
\end{center}
\end{table*}

%-------------------------------------
\vskip 6pt
{\bf Non-linearly realised symmetry.---}Expanding the unitary-gauge operators one further order in $\pi$ gives cubic operators for which all gradient-time mixed ones are uniquely fixed by the same coefficient that sets the quadratic mixing. This generalises the well-known fact that the operator $\dot\pi(\partial_i\pi)^2/a^2$ is fixed by the sound speed $c_\pi$ but not $\dot\pi^3$ in the single-field EFT. To find them, we use the following transformation law:
\begin{equation}
    (g^{0\mu}\partial_\mu\sigma_I) \to -\dot\sigma_I(1+\dot\pi) + \frac{(\partial_i\pi)(\partial_i\sigma_I)}{a^2} \,.
\end{equation}
We collect these operators in Tab.~\ref{tab: unitary-gauge operators for L3} where we also provide the exhaustive list of all other unitary-gauge operators that generate cubic interactions among fluctuations. Notice that restoring $e_2\neq0$ also necessarily fixes the operator $\dot\sigma_I(\partial_i\pi)(\partial_i\sigma_J)/a^2$ by symmetry.

%-------------------------------------
\vskip 6pt
{\bf Numerical implementation.---}The algebraic equation $\det \M(\Delta) = 0$ has no closed-form solution beyond $\Nf=3$, and a robust parameter scan requires solving it for at least $\O(10^6)$ points. The naive strategy---symbolically expand $\det \M(\Delta)$ into its scalar polynomial coefficients, then find its roots---is unsuitable because expanding an $(\Nf+1) \times (\Nf+1)$ symbolic determinant is itself combinatorially expensive and does not vectorise across parameter points. Root-finding directly from them is notoriously ill-conditioned at large $\Nf$: tiny coefficient perturbations can produce large, spurious shifts in the roots.

\vskip 6pt
{\it Eigenvalue problem.---}We instead exploit the fact that $\M(\Delta)$ is not an arbitrary scalar polynomial but a {\it matrix} polynomial of degree two,
\begin{equation}
    \M(\Delta) = A_0 + \Delta A_1 + \Delta^2 A_2 \,,
\end{equation}
with $A_0, A_1, A_2$ read off from~\eqref{eq: scaling-dimension matrix}. This is a quadratic eigenvalue problem, for which numerical linear algebra offers a standard and numerically robust tool. Introducing the auxiliary vector $\bar y\equiv\Delta\bar\chi$ turns the quadratic condition $\M(\Delta)\bar\chi = 0$ into the linear system
\begin{equation}
    \Delta
    \begin{pmatrix}
    \bar\chi \\
    \bar y
    \end{pmatrix} = 
    \begin{pmatrix}
    0 & I \\
    -A_2^{-1}A_0 & A_2^{-1} A_1
    \end{pmatrix}
    \begin{pmatrix}
    \bar\chi \\
    \bar y
    \end{pmatrix} \,,
\end{equation}
so that $\Delta$ is an eigenvalue of the $2(\Nf+1) \times 2(\Nf+1)$ matrix on the right-hand side, i.e.~exactly the $2(\Nf+1)$ roots of $\det \M(\Delta)=0$, now obtained from a single ordinary eigenvalue problem instead of a root-finding one. This requires $A_2$ to be invertible: comparing to~\eqref{eq: scaling-dimension matrix}, $A_2$ is exactly the kinetic matrix $\K_{AB}$ of~\eqref{eq: L2 all mixings}, which is positive definite, hence invertible, for any ghost-free theory, so the construction never fails. As a byproduct of computing $A_2^{-1}$, the no-ghost condition $\sum_I\eta_I^2<1$ and the effective mass positivity criterion~\eqref{eq: meff positivity} are evaluated at no extra cost: the former from a single norm, the latter directly from the real parts of the eigenvalues already computed.

\vskip 6pt
{\it Efficient solver.---}Every step of the problem (the matrix inverse, the two matrix products, and the eigenvalue decomposition) is applied identically to every parameter point in a scan. \texttt{numpy}'s linear algebra routines exploit this directly: \texttt{numpy.linalg.inv}, matrix multiplication, and \texttt{numpy.linalg.eigvals} all natively vectorise over leading array dimensions, so a batch of $\B$ parameter draws is processed as a single array of shape $(\B, 2(\Nf+1), 2(\Nf+1))$. The cost per parameter point is that of one $2(\Nf+1)\times2(\Nf+1)$ eigenvalue problem, and the whole batch is dispatched to compiled \texttt{LAPACK} routines at once. This way, the method is well-suited for scanning. For very large batches at large $\Nf$, the working array can exceed available memory before it exceeds available time. We then split the batch into smaller chunks processed sequentially, trading a controlled amount of Python-level looping for bounded peak memory, with no change to the underlying method or its accuracy.

\begin{figure}[h!] 
    \centering
    \includegraphics[width=0.9\linewidth]{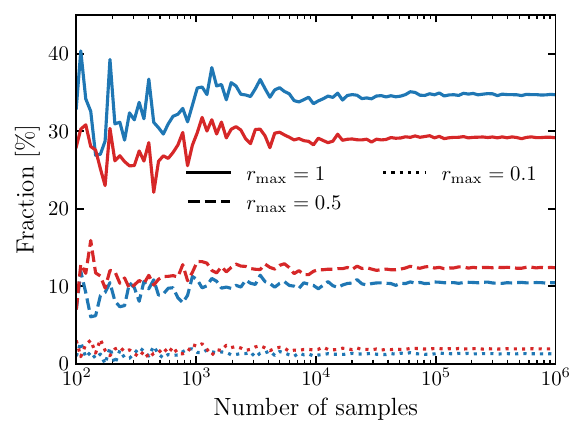}
    \caption{Fraction of the parameter space (in $\%$) satisfying $\Delta=r+i\mu$ with $\mu\geq0$ and $r\leq r_{\max}$ for $r_{\max}=1, 0.5$ and $0.1$, as a function of the number of drawn samples (for $\Nf=2$). The \textcolor{pyblue}{blue} lines correspond to the priors $m_I^2=[-1, 5]H^2, \rho_I=\omega=[0, 5]H$ and the \textcolor{pyred}{red} lines correspond to the priors $m_I^2=[-5, 10]H^2, \rho_I=\omega=[0, 10]H$. We have removed the $\Delta=0, 3$ massless modes.}
    \label{fig: convergence}
\end{figure}

\vskip 6pt
{\it Validation \& convergence.---}We validate this pipeline in two independent ways: against the $\Nf=2$ and $\Nf=3$ closed forms, and against the exact $\Delta=0,3$ roots, checked to hold at every sampled point to numerical precision. We have also checked that for $m_1^2=m_2^2=-(5H)^2, \rho_2=0.6H$ and $\omega=5H$ (all other parameters set to zero), we obtain $\Delta=0.06\pm5i$ (for the solution with the smallest real part), and that for $m_1^2=m_2^2=(0.3H)^2, \rho_2=0.4H$ and $\omega=H$ we obtain $\Delta= 0.46\pm i$, successfully recovering the results from~\cite{McAneny:2019epy}. In Fig.~\ref{fig: convergence}, we show the fraction of the parameter space which generates an enhanced squeezed bispectrum $\Delta=r+i\mu$ for $r\leq r_{\rm max}$ and $\mu\geq0$ for different sets of priors, setting $\Nf=2$. The fractions stabilise for $\sim 10^4$ samples, ensuring convergence. The prior for the scanned parameters does not play a significant role.

\begin{table}[h!]
    \centering
    \small
    \begin{tabular}{lccc}
    \toprule
    mass prior & $\kappa$ (at $\Nf=100$) & predicted $-\tfrac{1}{1+\kappa}$ & measured \\
    \hline\hline
    $\U(-10,10)$  & $-0.065$ & $-1.07$ & $-0.92$ \\
    $\N(0,1)$     & $-0.043$ & $-1.05$ & $-1.15$ \\
    $\N(0,5)$     & $-0.014$ & $-1.01$ & $-0.97$ \\
    $\N(0,10)$    & $-0.039$ & $-1.04$ & $-0.85$ \\
    $\N(1,1)$     & $+0.76$  & $-0.57$ & $-0.50$ \\
    \hline\hline
    \end{tabular}
    \caption{Predicted and measured power-law exponent of $\mathrm{median}(r_{\rm dom})$ as a function of $\Nf$. The predicted column uses only a single $\Nf=100$ fit to the pooled root density near $r=0$ (Eq.~\eqref{eq: prediction}). The measured column is the independent power-law fit to the full $\Nf=1$--$100$ scan.}
    \label{tab: predicted-vs-measured}
\end{table}

%-------------------------------------
\vskip 6pt
{\bf Solving $\det\M(\Delta)=0$ for $\Omega=0$.---}Writing $x\equiv \Delta(\Delta-3)$, $y\equiv 3-2\Delta$ and separating $\M(\Delta)$ into its $\pi_c$ row/column and the $\Nf\times\Nf$ $\sigma$-block $M(\Delta)\equiv \diag(x+(m_I/H)^2)+y\Omega/H$, the Schur complement of the $(1, 1)$ upper-left entry gives:
\begin{equation}
    \begin{aligned}
    \det\M(\Delta) &= x\det[M(\Delta)+\rho \rho^T/H^2] \\
    &= x\, \det M(\Delta)\left[1+\rho^T M(\Delta)^{-1}\rho/H^2\right]\,,
    \end{aligned}
\end{equation}
since $M(\Delta)+\rho\rho^T/H^2$ is a rank-one deformation of $M(\Delta)$. The first factor reproduces the exact $\Delta=0, 3$ roots trivially. All of the non-trivial physics is in the bracket. Writing $z\equiv -x$, this is precisely the secular equation
\begin{equation}
    \begin{aligned}
        &1-\rho^T[z-\H(\Delta)]^{-1}\rho=0 \,, \\
        &\text{with} \quad \H(\Delta) \equiv \diag(m_I/H)^2+y\,\Omega/H\,,
    \end{aligned}
\end{equation}
i.e.~the condition for $z$ to be an eigenvalue of $\H(\Delta)+\rho\rho^T$---a rank-one deformation of $\H(\Delta)$ by the mixing vector. Because $y=3-2\Delta$ also depends on $\Delta$, this is a genuinely implicit, self-consistent equation rather than a fixed eigenvalue problem. 

\begin{figure*}[t!]
  \centering
  \includegraphics[width=\linewidth]{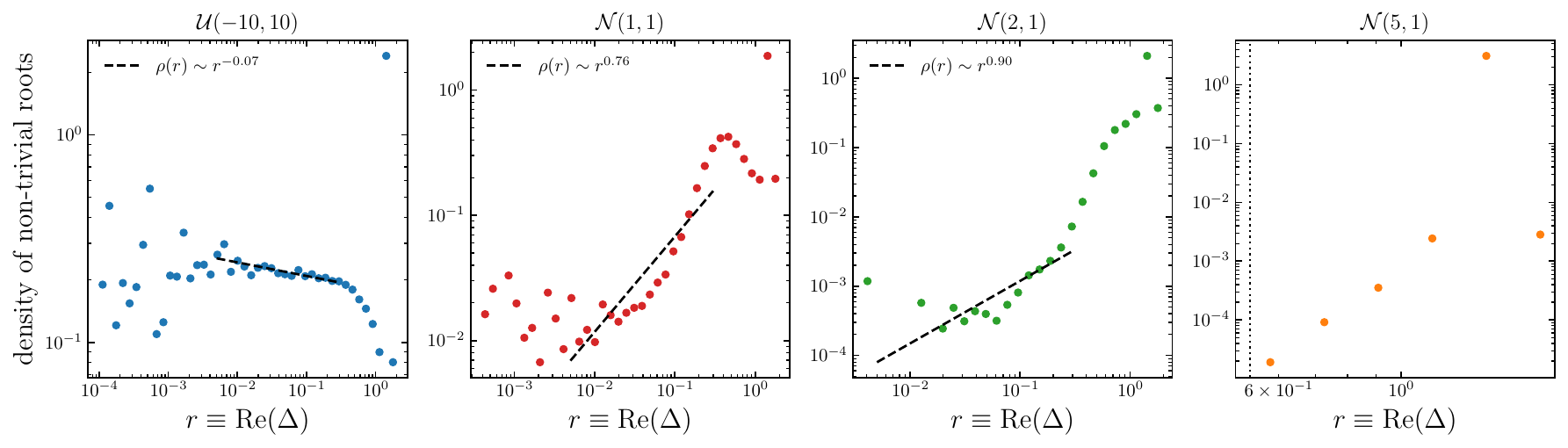}
  \caption{Pooled density of non-trivial root real parts near $r=0$ at fixed $\Nf=100$. {\it Left to right}: an essentially flat density ($\kappa\approx-0.07$, uniform mass), a depleted, rising density ($\kappa\approx+0.76$, $\N(1,1)$), a sparser density ($\kappa\approx+0.90$, $\N(2,1)$), and a genuine gap, with zero pooled roots below $r\approx0.5$ ($\N(5,1)$).}
  \label{fig: densityfits}
\end{figure*}

The case $\Omega=0$, where $\H$ collapses to the $\Delta$-independent diagonal mass matrix, is exactly solvable. Since $M(\Delta)$ becomes diagonal, the eigenvalue equation becomes the classic secular equation for a rank-one perturbation of a diagonal matrix
\begin{equation}
    f(z)=\sum_{I=1}^{\Nf} \frac{(\rho_I/H)^2}{z-(m_I/H)^2} = 1\,, \quad z=\Delta(3-\Delta) \,.
\end{equation}
Since the derivative $f'(z)=-\sum_{I=1}^{\Nf}(\rho_I/H)^2/[z-(m_I/H)^2]^2<0$ is negative everywhere, $f$ is strictly decreasing on each interval between consecutive poles. Near each pole from above, $f\to+\infty$, and from below we have $f\to-\infty$. Consequently, on each gap $(m_{(I)}^2, m_{(I+1)}^2)$ between sorted mass eigenvalues, $f$ sweeps monotonically from $+\infty$ to $-\infty$, crossing $f(z)=1$ exactly once. Below the smallest pole, every term is positive so $f(z)<0$ throughout. Above the largest pole, every term is negative and $f$ decreases from $+\infty$ to $0$, crossing $f(z)=1$ exactly once. We therefore obtain $\Nf$ real $z$-solutions, matching the polynomial's degree exactly. 

Each real $z$-root maps to $\Delta=3/2\pm \sqrt{9/4-z}$. It follows that complex modes are exactly pinned at $r=3/2$ for any mixing $\rho_I$ ($I=1, \ldots, \Nf$), which matches the exact localisation onto the principal series mechanism, observed in the main text (see Fig.~\ref{fig: localisation}). Beyond the $\Nf-1$ interlaced roots, the equation $f(z)=1$ always has exactly one further real root above the largest mass eigenvalue. Whether this specific root is parametrically separated from the bulk edge, rather than merging into it, is controlled by whether $\sum_I \rho_I^2/H^2$ exceeds a value set by the mass spectrum's edge. 

%-------------------------------------
\vskip 6pt
{\bf Random matrix theory for large-$\Nf$ ensemble.---}The drift of $r_{\rm dom}$ towards zero as $\Nf\to\infty$ follows from a single classical theorem for the minimum of many random quantities, applied to the root density of $\det\mathcal M(\Delta)=0$ near the origin.

\vskip 6pt
{\it General theorem.---}By construction, $r_{\rm dom}=\min_a\Re(\Delta_a)$ is the minimum of the $O(\Nf)$ non-trivial roots of $\det\mathcal M(\Delta)=0$ (the exact $\Delta=0,3$ roots of the $\pi_c$ sector excluded). Its large-$\Nf$ behaviour is therefore governed by extreme value theory, not by any feature specific to the model. Let $X_1,\dots,X_n$ be drawn i.i.d.~from a distribution with  finite lower endpoint $a$ and cumulative distribution function
\begin{equation}
\label{eq: cdf}
    F(a+\epsilon) \sim \epsilon^{\kappa+1}\,, \qquad \epsilon\to0^+\,,
\end{equation}
for some $\kappa>-1$ (the density need not vanish at $\epsilon=0$, may diverge mildly, or may be flat). The Fisher--Tippett--Gnedenko theorem then fixes~\cite{LRLRbook, Livan_2018}
\begin{equation}
\label{eq: min-scaling}
    \min(X_1,\ldots,X_n) - a \sim n^{-1/(1+\kappa)} \xrightarrow{n\to\infty} 0\,,
\end{equation}
for all $\kappa>-1$. This is the same law governing how the smallest eigenvalue of a random matrix approaches the edge of its spectrum.

\vskip 6pt
{\it Application to the ensemble.---}Let $g_{\Nf}(r)$ denote the (large-$\Nf$) density of non-trivial root real parts near $r=0$---$\Nf$-independent by construction, since the mass and $\Omega_{IJ}$ ensembles are drawn so as to hold the physical mixing scales fixed as $\Nf\to\infty$---and suppose $g_{\Nf}(r)\sim g_0\,r^{\kappa}$ as $r\to0^+$. Eq.~\eqref{eq: min-scaling} then predicts, with the number of candidate roots itself $O(\Nf)$,
\begin{equation}
\label{eq: prediction}
    \mathrm{median}(r_{\rm dom}) \sim \Nf^{-1/(1+\kappa)}\,.
\end{equation}
This holds for any mass and mixing prior. As long as the mass spectrum's support overlaps the scale set by $\rho_I,\Omega_{IJ}$, $g_{\Nf}(r)$ is finite and non-zero near $r=0$, $\kappa$ is finite, and Eq.~\eqref{eq: prediction} guarantees that $\mathrm{median}(r_{\rm dom})\to0$.

\vskip 6pt
{\it Numerical tests.---}We can test Eq.~\eqref{eq: prediction} by fitting $\kappa$ from the pooled root spectrum at a single, fixed $\Nf=100$, and comparing to the exponent independently measured by scanning $\Nf=1$--$100$. Tab.~\ref{tab: predicted-vs-measured} shows the results for five mass priors. All five agree with the parameter-free prediction to $\sim4$--$18\%$.

\end{document}